\documentclass[fleqn,10pt]{wlscirep}

\usepackage{bm, graphicx, color}
\usepackage{amsmath,amssymb,amsfonts}
\usepackage{ wasysym, hyperref}

\newcommand{\HH}{\mathcal{H}}
\newcommand{\PP}{P}

\title{Gravitational mass of positron from LEP synchrotron losses}

\author[1,2,*]{Tigran Kalaydzhyan}
\affil[1]{Department of Physics, University of Illinois, Chicago, Illinois 60607-7059, USA.}
\affil[2]{Department of Physics and Astronomy, Stony Brook University, Stony Brook, NY 11794, USA.}
\affil[*]{tigran@uic.edu}

\begin{abstract}
General relativity(GR) is the current description of gravity in modern physics. One of the cornerstones of GR, as well as Newton's theory of gravity, is the weak equivalence principle (WEP), stating that the trajectory of a freely falling test body is independent of its internal structure and composition. WEP is known to be valid for the normal matter with a high precision. However, due to the rarity of antimatter and weakness of the gravitational forces, the WEP has never been confirmed for antimatter. The current direct bounds on the ratio between the gravitational and inertial masses of the antihydrogen do not rule out a repulsive nature for the antimatter gravity. Here we establish an indirect bound of 0.13\% on the difference between the gravitational and inertial masses of the positron (antielectron) from the analysis of synchrotron losses at the Large Electron-Positron collider (LEP). This serves as a confirmation of the conventional gravitational properties of antimatter without common assumptions such as, e.g., coupling of gravity to virtual particles, dynamics of distant astrophysical sources and the nature of absolute gravitational potentials.
\end{abstract}

\begin{document}
\flushbottom
\maketitle

\thispagestyle{empty}
\section*{Introduction}

Since the formulation of the general theory of relativity (GR) by Einstein~\cite{Einstein:1915ca, Einstein:1916vd} in 1915, many efforts have been made to test the theory and no experimental contradiction has been found~\cite{Will:2014kxa}. One of the main statements in the foundation of GR (and even Newton's law of universal gravitation) is the equivalence of the gravitational and inertial masses, the so-called weak equivalence principle (WEP). WEP was successfully tested and proven at the $2\times 10^{-13}$ level for normal matter~\cite{Will:2014kxa}. However, since the theoretical prediction of antimatter by Dirac~\cite{Dirac:1928hu} in 1928 and its first experimental observation by Anderson~\cite{Anderson:1933mb} in 1933, there is no conclusive evidence of the gravitational properties of antimatter~\cite{Fischler:2008zz, Amole:2013gma}. This is despite the work with antiparticles becoming a commonplace practice in physics experiments and even medicine~\cite{PET}.

Direct observation of cold-trapped antihydrogen~\cite{Amole:2013gma} by the ALPHA collaboration at CERN sets the limits on the ratio between the gravitational, $m_g$, and inertial, $m$, masses of the antihydrogen, $-65 < m_g / m < 110$, including systematic errors, at the 5\% significance level. This ratio does not exclude the possibility of, e.g., repulsion of the antihydrogen by Earth (antigravity)! There are many indirect arguments against antigravity~\cite{Nieto:1991xq}. However, for the antimatter, all of them exploit some additional assumptions, such as gravitational properties of virtual particles, physical significance of the absolute values of the gravitational potentials, CPT-invariance, etc. One of the cleanest indirect tests is the comparison of decay parameters of the kaon-antikaon system in the presence of periodic (annual, monthly and diurnal) gravitational potential variations~\cite{Apostolakis:1999tk}. The equality between the kaon and antikaon gravitational masses was established at the $1.8\times 10^{-9}$ level~\cite{Apostolakis:1999tk}. However, the gravitational properties of the kaon itself and any other strange matter are not known.

In this article, we constrain possible violations of WEP for antimatter (more precisely, positrons) by analyzing synchrotron losses for 80~GeV positrons at the Large Electron-Positron collider (LEP) at CERN. The advantage of the accelerator experiments is the large relativistic $\gamma$-factor. The large $\gamma$-factor is known to reveal possible Lorentz-violating effects~\cite{Altschul:2009xh, Kalaydzhyan:2015ija, Kalaydzhyan:2015kxa} and to suppress electromagnetic interactions~\cite{Chao:1993zn}, which otherwise overwhelm gravitational forces~\cite{fairbank}. 
In addition, the accelerator experiments (in comparison to astrophysical observations) do not require additional and somewhat controversial assumptions on the dynamics of high-energy sources~\cite{Pakvasa:1988gd}. Analysis presented in this article, obviously, does not diminish the importance of the direct methods, since it is by itself indirect and, hence, model-dependent. 

\section*{Theory in brief} 
In the absence of gravity, a positron with charge $e$, velocity $\textbf{v}$, energy ${\cal E}$, inertial rest mass $m_e$ and acceleration $\dot{\textbf{v}}$ perpendicular to $\textbf{v}$
looses its energy through the synchrotron radiation with the power~\cite{Jackson}
\begin{align}
\PP = \frac{2}{3}\frac{e^2 \dot{\textbf{v}}^2}{c^3} \left(\frac{{\cal E}}{m_e c^2}\right)^4\,,\label{synch1}
\end{align}
where 
$c$ is the speed of light (in what follows we will work in the natural units, $c = \hbar = 1$). The case $\dot{\textbf{v}}\bot \textbf{v}$ corresponds to, e.g., synchrotron radiation in a constant homogeneous magnetic field.

The gravitational field of the Earth (Sun or other distant massive celestial objects) around the accelerator can be considered homogeneous and described by an isotropic metric for a static weak field, 
\begin{align}
ds^2 = \HH^2 dt^2 - \HH^{-2}(dx^2 + dy^2 + dz^2)\,,\label{metric}
\end{align}
where $\HH^2 = 1+2\Phi$, and $\Phi$ is the gravitational potential, defining the acceleration of free-falling bodies.
For a massive particle (in our case positron) with gravitational mass $m_{e,g}$, one can write the gravitational potential as~\cite{Kalaydzhyan:2015ija}
\begin{align}
\Phi_m = \Phi \,\frac{{m_{e,g}}}{m_e}\,, \quad \qquad \HH_m^2 \equiv 1+ 2\Phi_m\,,\label{modif}
\end{align}
which will modify the dispersion relation of the positron with momentum $\textbf{p}$ and energy $\mathcal{E} \gg m_e$ and the relation between energy and mass~\cite{Kalaydzhyan:2015ija},
\begin{align}
\textbf{p}^2 = (1-2\kappa) \left(\mathcal{E}^2 - m_e^2 \right),\quad\qquad
{\cal E} = \frac{m_e \HH^{-1}\HH_m}{\sqrt{1-\HH^{4}\HH_m^{-4}\textbf{v}^2}}, \label{energy}
\end{align}
where $\kappa = 2 \Phi \Delta m_e / m_e$, $\Delta m_e = m_{e,g} - m_e$. We consider no change in the photon dispersion relation due to strong constraints on the variation of the speed of light~\cite{Ackermann:2009aa, Biller:1998hg, Schaefer:1998zg}. Parameter $\kappa$ here plays a role of an anomalous redshift (or blueshift) and vanishes in the limit $m_{e,g}\rightarrow m_e$. We also assume $|\kappa| \ll 1/\gamma^2$, which is the case at LEP, as will be seen later. This modification can be described by the isotropic version of Standard Model Extension~\cite{Colladay:1996iz, Colladay:1998fq, Kostelecky:2003fs} (SME) with $c_{00}=3c_{ii}=3\kappa /4$ and other Lorentz-violating parameters set to zero. Eq.~(\ref{modif}) is a way to generalize the gravitational coupling of a massive test particle to the background which reproduces the Newton's gravitational law and its relativistic extension. Other modifications of the metric and new accelerator phenomenology emerging from them are to be studied elsewhere. [For a more general form of the background metric, $ds^2 = \Sigma^2 dt^2 - \Xi^{-2}(dx^2 + dy^2 + dz^2)$ and corresponding metric coefficients $\Sigma_m$ and $\Xi_m$ for a massive particle, the Eqs.~(\ref{energy}) become $\textbf{p}^2 = \left( \frac{\Xi}{\Xi_m}\right)^2 \left[ \mathcal{E}^2  \left( \frac{\Sigma}{\Sigma_m}\right)^2 - m_e^2\right]$ and $\mathcal{E} = m_e \left( \frac{\Sigma_m}{\Sigma}\right)\left[ 1- \left( \frac{\Xi\,\, \Sigma}{\Xi_m \Sigma_m}\right)^2 \textbf{v}^2\right]^{-1/2}$]. Our analysis should be considered preliminary in the sense that we do not study the cases of WEP violation that cannot be incorporated in the change of the Newton's constant or difference between the gravitational and inertial masses.

In an ultrarelativistic case, $|\textbf{v}| \approx 1$, the ratio ${\cal E}/m_e$ from Eq.~(\ref{energy}) 
modifies the synchrotron radiation power (\ref{synch1}) by an amount $\Delta \PP$, such that
\begin{align}
\Delta \PP / \PP  = 4 \kappa \gamma^2 \,,\label{synch2}
\end{align}
where $\gamma \equiv 1/\sqrt{1-\textbf{v}^2}$. This (naive) derivation leads to the same result as a much more rigorous analysis of the synchrotron radiation within the SME~\cite{Altschul:2005za, Altschul:2009xh}. Now, let us imagine that two sets of experimental data ``1'' and ``2'' will restrict the values of $\kappa$ by $|\kappa|< \kappa_{1,2}\equiv|\Delta P/P|_{1,2}/(4 \gamma^2)$
at the moments when the gravitational potential acquires values $\Phi_1$ and $\Phi_2=\Phi_1+\Delta \Phi$, respectively. The difference $\Delta \Phi$ can be related to the periodic variations in the distances between Earth and other celestial bodies. The experiments are assumed to reproduce the conventional synchrotron radiation power within the uncertainties $(\Delta \PP / \PP)_{1,2}$ determining $\kappa_{1,2}$.
The restriction on the gravitational mass is then given by
\begin{align}
\left|\frac{\Delta m_{e}}{m_e}\right| < \frac{\kappa_1 + \kappa_2}{2 |\Delta \Phi|}\,.\label{limit1}
\end{align}
On the scale of several months, one can take the change in the solar potential on the Earth's surface as the leading contribution to $\Delta \Phi$. Due to the eccentricity of the Earth's orbit, the distance between Earth and Sun, $d_{SE}\approx 1$~AU (astronomical unit), varies by the amount $\Delta d_{SE} \ll d_{SE}$. This changes the solar potential, $\Phi_\odot = -G_N M_\odot/d_{SE}\approx -9.9\times 10^{-10}$, by $\Delta\Phi = - \Phi_\odot \Delta d_{SE} / d_{SE}$, where $G_N$ is the Newton's constant and $M_\odot$ is the solar mass.
Substituting this into (\ref{limit1}), we obtain the relation between the fractional deviation in the masses and the fractional uncertainty in the measured synchrotron radiation power in two experiments,
\begin{align}
\left|\frac{\Delta m_{e}}{m_e}\right| < \frac{|\Delta \PP / \PP|_{1} + |\Delta \PP / \PP|_{2}}{8 \gamma^2 |\Phi_\odot \Delta d_{SE}[\mathrm{AU}]|}\,,\label{limit2}
\end{align}
where we also assumed the $\gamma$-factor to be the same in both experiments.

\section*{Analysis of the experiment} 
In order to apply the formula (\ref{limit2}), we consider an analysis of the LEP Energy Working Group~\cite{Assmann:2004gc} for the LEP 2 programme in the last few years of LEP operation. The primary physics motivation for the LEP 2 programme was precise determination of the W boson mass, $M_\mathrm{W}\approx 80.4$~GeV. For this purpose, the relative error of the centre-of-mass energy for the accelerated electrons and positrons was reduced to $\Delta E_{\mathrm{CM}} / E_{\mathrm{CM}} = 1.2 \times 10^{-4}$ for the majority of the LEP 2 runs~\cite{Assmann:2004gc}. There were three complementary approaches used by the group: precise measurement of the bending magnetic field by nuclear magnetic resonance (NMR) probes and the flux-loop~\cite{Assmann:2004gc}; spectrometry, i.e. beam deflection by a precisely known magnetic field in a bending magnet; and analysis of the synchrotron tune as a function of the beam energy, energy loss per turn and the accelerating radiofrequency (RF) voltage. A full description of the methods and results is given in the Report~\cite{Assmann:2004gc}; the LEP machine description can be found in the technical design reports~\cite{LEPv1, LEPv2, LEPv3}.

The first two methods would not be affected by the Lorentz-violation, since the magnetic field modification is ruled out by the atomic clock experiments~\cite{Mattingly:2005re}. The third method is based on the measurement of the synchrotron tune, $Q_s$, which is defined as a ratio of the frequency of longitudinal fluctuations in the beam (due to the synchrotron radiation loss and boost from the accelerating RF system) to the revolution frequency. The fitting formula for $Q_s$ used for the determination of the beam energy $E$ was~\cite{Assmann:2004gc}
\begin{align}
Q_s^4 = \left(\frac{\alpha_c h}{2\pi}\right)^2\left\{ \frac{g^2 e^2 V_{RF}^2}{E^2}+ M g^4 V_{RF}^4 - \frac{\tilde U^2}{E^2}\right\}\,,\label{fit}
\end{align}
where $\alpha_c$ is the momentum compaction factor (measure for the change in the orbit length with momentum), $h$ is the harmonic number of the accelerator (ratio between the RF frequency and the revolution frequency), $V_{RF}$ is the amplitude of RF voltage, $g \approx 1$ is a fitting parameter whose role is to account for uncertainties in the accelerating RF system, $M \sim 10^{-7}$ is a parameter accounting for the fact that the RF voltage is not distributed homogeneously around the accelerator but in four straight sections near to even numbered access points~\cite{LEPv3} (parameter is obtained from simulations) and $\tilde U$ is the total energy loss per turn (synchrotron radiation in dipole magnets, quadrupole magnets, parasitic mode losses, etc., either modeled or measured directly). The total $1\sigma$ error of this method, $\Delta E$, as well as the difference between the fitted energy ($E=80$~GeV) and the NMR model, $E^{Q_s}-E^{\mathrm{NMR}}$, for several LEP runs are shown in Table~\ref{fills}.

Following Altschul~\cite{Altschul:2009xh}, we notice that since $E$ was more accurately known from the other complementary methods, one can reinterpret Eq.~(\ref{fit}) as a fit to $\tilde U$ (which we, for conservative estimates, consider to be dominated by the radiation in dipoles and, therefore, proportional to $\PP$). Treating $E$ as an exact quantity, the previously estimated relative error for $E$ becomes a relative error for $\tilde U$ and, hence, for $\PP$.

\section*{Results} 
Choosing two sets of measurements (fills 6114 and 6338) and taking into account the difference between NMR and $Q_s$ values for energies, we obtain $|\Delta \PP / \PP|_{1,2} < 9\times 10^{-4}$ as a $2\sigma$ bound for both measurements, which with use of Eq.~(\ref{limit2}) finally translates to
\begin{align}
\left|\frac{\Delta m_{e}}{m_e}\right| < 1.3\times 10^{-3}\,,\label{limit3}
\end{align}
i.e. a 0.13\% limit on a possible deviation from WEP. This significantly improves our previous 4\% limit~\cite{Kalaydzhyan:2015ija} coming from the absence of the vacuum Cherenkov radiation and photon stability in accelerator experiments. Since the energy calibration at LEP was performed for both electrons and positrons, the same limit is applied for the electrons. Also, note that $|\kappa_{1,2}|<9.2\times 10^{-15} \ll 1/\gamma^2 = 4.1\times 10^{-11}$, so the previously made assumption is justified.

The weak equivalence principle was introduced in some form already in the 17th century by Newton in his \textit{Principia}. It was further employed by Einstein in the 1910's. Furthermore, the antimatter was discovered in the 1930's. However, only now, with the advances in accelerator physics, are we finally capable to draw certain conclusions on how to bring together the concepts of gravity and antimatter.


\section*{Acknowledgements}
This work was supported in part by the U.S. Department of Energy under Contracts No. DE-FG-88ER40388 and DE-FG0201ER41195. The author would like to thank Dmitry Duev for astronomical references.
\section*{Competing Interests}
The author declares that he has no competing financial interests.

\begin{table}
\centering
\caption{Variations in the distance between Earth and Sun during some of the $E=80$~GeV LEP runs together with the total error in the energy determination assigned by the LEP Energy Working Group~\cite{Assmann:2004gc}. The distances are obtained from the NASA's Jet Propulsion Laboratory (JPL) solar system data~\cite{jpl}\label{fills}}
\medskip
\begin{tabular}{ccccc}
\hline
Fill & ~~~Date~~~ & ~~~$d_{SE}$~~~  & $E^{Q_s}- E^{\mathrm{NMR}}$& $\Delta E$ \\
number & of the fill & [AU] & [MeV] & [MeV]\\
\hline
6114 & 13 Aug 1999 & 1.013 & -4 & 30\\
6338 & 15 Sep 1999 & 1.006 & 10 & 30\\
8315 & 29 Aug 2000 & 1.010 & -10 & 62\\
8445 & 10 Sep 2000 & 1.007  & -52 & 38\\
8809 & 18 Oct 2000 & 0.996 & -43 & 52\\
\hline
\end{tabular}
\end{table}

\end{document}